\documentstyle[twocolumn,aps]{revtex}
\input psfig.sty

\def\ltsim{\vbox {\hbox{\lower 0.6\baselineskip \hbox{$<$}} \break
		 \hbox{\lower 0.1\baselineskip \hbox{$\sim$}} } }
\def\gtsim{\vbox {\hbox{\lower 0.6\baselineskip \hbox{$>$}} \break
                 \hbox{\lower 0.1\baselineskip \hbox{$\sim$}} } }
\def\k{{\bf k}}

\def\R{{\bf R}}

\def\vs{{\bf v}_s}
\begin{document}
\draft

\twocolumn[\hsize\textwidth\columnwidth\hsize\csname %
@twocolumnfalse\endcsname

\title{Transport Properties of  $d$-Wave Superconductors in the Vortex State}

\author{
C. K\"ubert and P.J. Hirschfeld
}

\address{
Department of Physics, University of Florida, Gainesville, FL 32611, 
USA.\\
}

\maketitle
\begin{abstract}
We calculate the magnetic field dependence of quasiparticle transport 
properties in the vortex state of a $d$-wave superconductor arising solely 
from the quasiparticle's Doppler shift in the superflow field surrounding 
the vortex. Qualitative features agree well with experiments on cuprate 
and heavy fermion superconductors at low fields and temperatures. 
We derive scaling relations in the variable $T/H^{1/2}$ valid at sufficiently 
low temperatures $T$ and fields $H$, but show that these relations depend on 
the scattering phase shift, and are in general fulfilled only approximately 
even in the clean limit, due to the energy dependence of the quasiparticle 
relaxation time. 
\end{abstract}
]
{\it Introduction.} 
While the origins of high-temperature superconductivity  and the  non-Fermi
liquid nature of the normal state in the cuprates are not understood, recently a consensus
has been emerging that the superconducting state itself is not particularly exotic, in the
sense that it consists of a BCS-like pair  state with well-defined quasiparticle excitations
above it.  Although the order parameter in these materials is now  thought to display
$d$-wave symmetry,
ranking them  among the few known ``unconventional"
superconductors,  it has been argued that the prediction of observable properties in the
superconducting state is now a relatively routine task. 


  In the study of transport
properties, however, questions arise which prevent one from drawing such sanguine
conclusions. Recent measurements of thermal conductivity in
both the LSCO-214 and BSCCO-2212 systems have hinted at qualitatively new physics in the
vortex state at low temperatures and fields.\cite{Ong214,Ong2212}  In the LSCO
case, a logarithmic dependence on field has been found suggestive of quantum
interference effects; in the BSCCO case, a kink in the thermal conductivity as a
function of field occurs, apparently signaling a transition to a new
superconducting phase at higher fields. 

In order to extract which aspects of the new phenomena are due to qualitatively ``new"
physics, one needs first to understand which aspects can be
attributed to the perhaps more mundane but still largely unexplored phenomenology of
quasiparticles in the $d$-wave vortex state.  A fundamental observation regarding
the thermodynamics of this state was made by Volovik, who pointed out that, in 
contrast to classic superconductors, in $d$-wave systems the entropy and density of states
was dominated by contributions from extended quasiparticle states rather than the
bound states associated with vortex cores.  The remarkable consequence of this
observation, which occurs due to the existence of order parameter nodes in the
clean $d$-wave state, is a term in the specific heat of the superconductor
in a field varying as
$\sqrt{H}T$ rather than as
$HT$ as in the classic case.  Other  calculations, dependent only on the
 Dirac-like spectrum of nodal quasiparticles excited at low temperature, gave the
field dependence of the density of states $N(\omega;H)$\cite{KopninVolovik1}
and predicted the specific heat 
 should scale as $C(T;H)/T^2\sim F_C(X)$, where $X\equiv
T/H^{1/2}$.\cite{SimonLee,KopninVolovik2}  Similar scaling behavior was predicted for
transport properties. Experimentally,
a $\sqrt{H}T$ term has indeed been identified in specific heat measurements on single crystals
of $YBCO$,\cite{Moleretal} but other measurements find deviations from this
behavior.\cite{Fisheretal,Revazetal}   Scaling behavior has been reported in the specific
heat of $YBCO$,\cite{Junodscaling} but with significant scatter  over a limited range.
Very recently, impressive scaling results for the thermal conductivity of single crystals
of the unconventional heavy fermion superconductor $UPt_3$ were also obtained by Suderow et
al.\cite{Flouquetscaling}

In this work we argue that  transport at low
magnetic fields and temperatures in  superconductors with line 
nodes in the gap  
can be understood within a model in which quasiparticles are
scattered by the same physical processes which dominate transport in the zero-field state. 
Magnetic field dependence of transport coefficients arises because of changes in quasiparticle
occupation numbers due to their Doppler shifts in the presence of the superfluid flow
around vortices, and can therefore be calculated from the energy dependence of the
zero-field relaxation time.  We argue that scattering by vortex cores, which in classic
superconductors usually determines the quasiparticle 
mean free path, is probably negligible
in $UPt_3$ and high-temperature superconductors (HTSC) over a suprisingly wide range of $H$ and $T$.  

The current approach is 
based on a semiclassical treatment of the extended quasiparticle states, including the
effects of impurity scattering.\cite{KH}  In an attempt to explain some of the
discrepancies in specific heat measurements,\cite{KH} 
we showed that the existence of a
finite impurity bandwidth
$\gamma$ destroys the scaling properties expected in the clean limit when $\gamma$ becomes
comparable to the average quasiparticle Doppler shift, $E_H=a(H/H_{c2})^{1/2}\Delta_0$, where
$H_{c2}$ is the upper critical field, $\Delta_0$ is the gap maximum over the Fermi surface,
and $a$ is a vortex-lattice dependent constant of order unity.  
Results for $N(0)$ in the dirty limit similar to ours were obtained
independently by Barash et al.\cite{Barashetal}

In this Letter we calculate transport properties in d-wave
superconductors using the formalism developed in Ref.
\cite{KH}. We find that the  low-temperature conductivities {\it increase} as a function
of field $H$ as $\sigma,\kappa^{el}/T\sim H \log H$ for unitarity limit
scatterers, in consequence of the excess occupation of extended quasiparticle
states induced by the field.  With increasing temperature, the transport coefficients
cross over to logarithmically {\it decreasing} dependence on $H$ resulting from the
 energy dependence of the unitarity limit scattering amplitude.   
 In contrast to the predictions of Simon and Lee\cite{SimonLee}, we find that exact 
one-parameter scaling with $X=T/H^{1/2}$ is not necessarily obtained.    As in Refs.
\cite{Barashetal,KH}, when the impurity bandwidth $\gamma$ is comparable to either the
temperature T or the magnetic energy
$E_H$, the simple scaling laws do not apply (although 2-parameter scaling relations may be
derived).  However, even in the clean limit $\gamma\rightarrow 0$, pair correlations
induce nonanalytic terms in the quasiparticle relaxation time which weakly violate
scaling.  Effects of this kind appear to have been neglected in Ref.
\cite{SimonLee}.  However,  approximate scaling functions may nevertheless be derived
whose form depends on the model of the quasiparticle relaxation time chosen.  Observations
of scaling in transport properties can therefore provide microscopic information on
the quasiparticle scattering amplitude. 

{\it Validity of the model.}
The approximation adopted is a simple generalization
to finite fields of the ``dirty d-wave theory" which has been quite successful in
describing properties of the cuprates and $UPt_3$.    The motion of quasiparticles is
determined by their energy in the frame of the moving superfluid,  assumed
for low fields to be well approximated by the single-vortex superfluid velocity
$\vs=\hbar{\hat
\theta}/ 2mr$, valid up to a cutoff of order ${\rm min}\{R,\lambda\}$, 
where $2R=\sqrt{2\pi}a^{-2}H_{c2}/H$ is the
intervortex spacing, $a$ is a lattice geometrical factor of order unity, and
$\lambda$ is the penetration depth.  The primacy of the extended quasiparticle
states and their description in terms of a propagator formalism for the calculation
of thermodynamic properties have been discussed in Refs.
\cite{Volovik1,KH}.   What is less clear is that such a method can be used for the
description of transport properties, normally described in classic superconductors 
in terms of a mean free time for the scattering of quasiparticles by vortices.
While such processes undoubtedly occur, we take here the point of view that at
sufficiently
low temperatures and fields, the
quasiparticle mean free path $\ell$ is determined by impurity (and 
electron-electron) scattering rather than by vortex scattering.  

Determining over how wide a
range of fields and temperatures this assumption applies is a difficult problem, as the
nature of quasiparticle scattering from a
$d$-wave vortex is not completely  understood. One must in general simultaneously account
for the phase shift of the quasiparticle wave function by the superfluid flow field around
each vortex, and for the scattering by Andreev reflection at the
core.\cite{SaulsRainerAndreev}  
We content ourselves in this Letter
with the  observation that the quasiparticle mean free path 
$\ell_{vortex}$ due to vortex scattering alone
is at low temperatures considerably longer than the intervortex spacing $R$,
since it arises solely from the disorder-induced or
thermally excited fluctuations of the ideal periodic Abrikosov lattice.
Furthermore, the bare potential for scattering derives
ultimately from the Coulomb interaction between the scattered 
quasiparticle and the bound
states in the vortex core; since there 
are many fewer such states in short coherence length systems, the
potential is correspondingly weaker and mean free paths
longer than in classic
superconductors. 

{\it Green's function for extended quasiparticle states.}  
 In the semiclassical approach one assumes a spatially constant order parameter $\Delta_k$,
thereby neglecting vortex core bound states, whose contribution to the density of states
may be shown in the $d$-wave case to be negligible.\cite{Volovik1}  These states do
participate in carrying a supercurrent through the core via their overlap with the
extended states,\cite{SaulsRainerAndreev} but may be neglected in the calculation
of bulk transport properties.  For concreteness, we 
assume here a system characterized by zero
field  $d_{x^2-y^2}$ equilibrium 
order parameter modelled by  the usual simplified form
$\Delta_k=\Delta_0\cos 2\phi$ over a 2D circular Fermi surface. 
In the vortex state quasiparticle energies are Doppler shifted by $\delta
\omega_\k(\R)=\vs\cdot\k$.  

With these assumptions, the single-particle matrix Green's function
in the presence of impurity scattering is\cite{KH}
\begin{eqnarray}
g(\k,\omega) = \frac
{  ( \tilde\omega-\vs\cdot\k) \tau_0 +  \Delta_{\k}\tau_1 
                         +  \xi_{\k} \tau_3 }
{(\tilde\omega-\vs\cdot\k)^2 -  \Delta_\k^2 -\xi_\k^2} \; ,
\end{eqnarray} where the $\tau_i$ are  Pauli matrices in particle-hole space.
In (1),  $\xi_k$ is the usual single-particle band measured relative to
the
Fermi level.  The renormalized frequency $\tilde\omega=\omega-\Sigma_0$ is position-dependent
due to the self-consistent coupling to the flow field in the usual t-matrix approximation
for impurity scattering,\cite{HVW,SMV} $\Sigma_0(\omega)=\Gamma G_0 \tau_0/( c^2-G_0^2 )$,
where $\Gamma=n_i/\pi N_0$ is an impurity scattering rate proportional
to  the concentration $n_i$ of point potential scatterers, 
$c=\cot \delta_0$ is the cotangent of the s-wave scattering phase shift
$\delta_0$, and $N_0$ is the density of states at the Fermi level.
  We will be primarily interested here in the strong scattering limit 
$\delta_0\simeq \pi/2$, as demanded by experimental results in zero magnetic field, but
results are easily obtained for arbitrary phase shift.
The averaged integrated Green's function is 
$G_0(\omega) = (\pi N_0)^{-1}\Sigma_{\k} {1\over 2}\;{\rm Tr}\;\tau_0\;
g(\k,\omega) $
We note  that  
the most general form of the propagator would
include renormalizations of both the single-particle energy $\xi_\k$ and
the order parameter $\Delta_k$ as well, but these may be safely 
neglected.\cite{KH}

{\it Electronic thermal conductivity in  magnetic field }.
Application of the Kubo formula to the case of a homogeneous temperature gradient gives
the following expression for the normalized electronic 
thermal conductivity 
(the $\omega\rightarrow 0$ electrical conductivity $\sigma$ is similar),
\begin{eqnarray}
\label{thermal conductivity}
\frac{\kappa^{el}_i(T;\R)/T}{\kappa^{el}_{i}(T_c)/T_c}
&=&
\frac{6}{\pi^2}
\int_0^{\infty}{d\omega}\left(\frac{\omega}{T}
\right)^2 \left( \frac{-\partial f}{\partial\omega}\right)
K_i(z , T)
\\
\label{K}
K_i(\tilde\omega, T)
 &=&
\frac{\Gamma^{tot}}{\tilde{\omega}^\prime
\tilde{\omega}^{\prime\prime}} \:
{\rm Re} 
\left< \hat k_i^2 \:
\frac{\tilde{ \omega}^2 + |\tilde{\omega}|^2 - 2|\Delta_{\bf k}|^2}
     {\sqrt{\tilde{\omega}^2 - \Delta_{\bf k}^2}} 
\right>
\end{eqnarray}
\noindent
where the Doppler shifted renormalized frequency is $z\equiv \tilde\omega-\vs\cdot\k$, and
$\Gamma^{tot}$ is the sum of the elastic and inelastic scattering rates at $T_c$.
 This
represents the contribution to the heat current from a small region of space around $\R$, in
which the local superfluid  velocity is $\vs(\R)$,
from quasiparticles at the Fermi 
level.  Impurity vertex corrections, which vanish
identically at 
$H=0$,\cite{HWE} are still negligible, of 
order $H\gamma^2/(H_{c2}\Delta_0^2)$.  Here 
$\gamma\equiv\Sigma_0^{\prime\prime}(0)$ is the broadening of
the nodal quasiparticle states.
  

Evaluation of Eqs. (2-3) is straightforward but numerically intensive due to the
coupling of momentum and real space degrees of freedom.  In the clean limit this coupling
may be shown to vanish, which is equivalent to replacing $\vs\cdot \k$ with $\vs\cdot \k_n$,
where $\k_n$ is a nodal wave vector, and symmetrizing over the four $d_{x^2-y^2}$ nodes.
In this case the momentum integrals may be performed analytically, simplifying the treatment
considerably. Even for finite impurity concentrations, where this replacement is
not exact, it provides an excellent approximation,as shown in \cite{KH}.  Note this is not
equivalent to a complete linearization of the system about the nodes; corrections involving
the gap scale $\Delta_0$ are retained in order to discuss deviations from pure Dirac scaling.

{\it Clean limit.}
 We first give results for the in-plane thermal conductivity at
$T=0$ in the clean limit,
where the impurity scale $\gamma$ is
much smaller  than the magnetic field energy $E_H$. In this case
\begin{eqnarray}
{\delta \kappa^{el}(\R )\over \kappa_{00}} = 
\frac{\pi}{4}\frac{\Delta_{0}}{\Gamma}
\left( \frac{\vec v_s \cdot \vec k_n}{\Delta_0} \right)^2 ,
\end{eqnarray}
where $\kappa_{00}/T\equiv \pi
N_0v_F^2/(6\Delta_0)$ is the universal
$d$-wave thermal conductivity,\cite{PALee,TCuniversal} with $v_F$ the Fermi
velocity.  Recently the magnitude of $\kappa_{00}$ and its independence of 
disorder was confirmed for $YBCO$ single crystals by Taillefer et al.\cite{Tailleferuniversal} This expression needs to be averaged over  a vortex to give
the field dependence.  If the current ${\bf j}_Q$ is
$\parallel$ to the field $\bf H$, it is clearly  appropriate to
average currents from each part of the vortex in parallel, giving the
naive average $\langle
\kappa\rangle_{H,\parallel}\equiv  A^{-1}(H)\int_{\rm cell} d^2r\ \kappa (\R)$,
where the vortex lattice unit cell area is $A\simeq \pi R^2$.   On the other hand, for
configurations with ${\bf j}_Q \perp {\bf H}$, a given current density element samples
many regions of different conductivity, so a series average is appropriate, $\langle
\kappa\rangle_{H,\perp}\equiv (\langle \kappa^{-1}\rangle_{H,\parallel} )^{-1}$.
At $T=0$  the field dependence for $H\ll H_{c2}$ 
is then
\begin{eqnarray}
{\kappa^{el}(H)\over  \kappa_{00}}=\left \{ \begin{array}{ll}
1+ s ({H}/{H_{c2}}) \:
\ln   {\pi\over 2 a^2} {{H_{c2}}\over{H}} &{\bf j}_Q \parallel {\bf H}\\
\rho^2/(\rho\sqrt{1+\rho^2}-\sinh ^{-1} \rho)& {\bf j}_Q \perp {\bf H}
\end{array}
\right.,
\end{eqnarray}
with $s= a^2\pi^2 \Delta_0/(16 \Gamma)$, $\rho \equiv [(8 \Gamma H_{c2})/(\pi^2 a^2 \Delta_0
H)]^{1/2}$.
 We note an unexpected result, that in both cases the
application of a field {\it increases} the thermal conductivity in this picture, due
to the extra current-carrying quasiparticle  states created near the nodes by the field.
This is  contrary to the usual picture, in which increasing the density of
vortices  leads to a 
reduction in the mean free path and hence a smaller conductivity.
Although in the present theory 
  $\kappa^{el}(T\rightarrow 0;H=0)=\kappa_{00}$ 
is universal in the sense of Lee,\cite{PALee,TCuniversal} the field
corrections (5) are {\it not}.
\begin{figure}[h]
\begin{picture}(150,260)
\leavevmode\centering\includegraphics{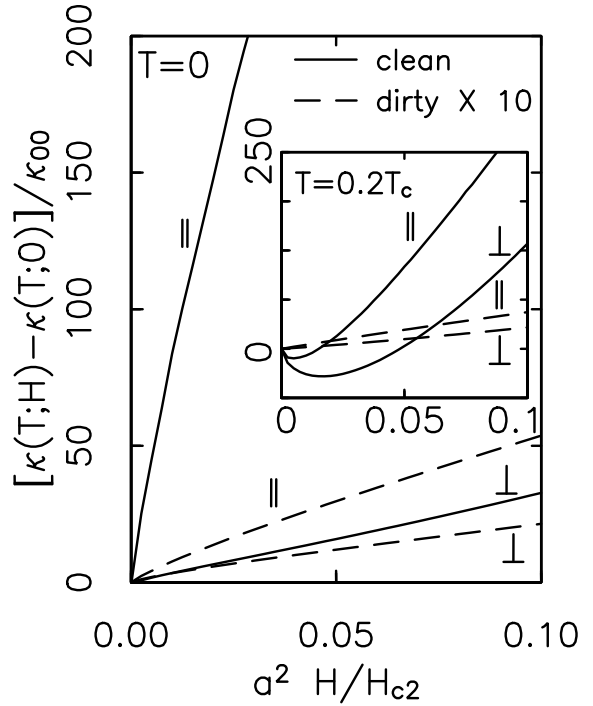}
\end{picture}
\caption{  
Magnetic field dependence of thermal conductivity with unitarity limit scattering for
$T=0$ and 
$\Delta_0/T_c=2.14$.  Solid lines: $\Gamma/T_c=0.001$;  dashed 
lines: 10 $\times$ $\delta\kappa/\kappa_{00}$ for $\Gamma/T_c=0.1$.  Symbols $\parallel$ and $\perp$ indicate 
averages with $j_{\bf Q}\parallel {\bf H} $ and $j_{\bf Q}\perp {\bf H} $,
respectively.  Insert: same for $T=0.2 T_c$. }
\end{figure}
At finite temperatures and fields such that $0<\gamma<T,E_H\ll\Delta_0$, 
 the following
leading log approximation may be obtained for the case ${\bf j}_Q \parallel {\bf H}$,
\begin{eqnarray}
\frac{\kappa^{el}(T;H)-\kappa^{el}(H)}{\kappa_{00}}&=&{7\pi 
T^2\over 5\Delta_0\Gamma}
\left(\ln^2 {4\Delta_0\over 3.5T+E_H} +
{\pi^2\over 4}\right). 
\end{eqnarray}
The expression (6) includes the correct asymptotic limits for
$\kappa(T)$ at  $H=0$ and $\kappa(H)$ at $T=0$.    In Fig. 1,
we exhibit the  full numerically determined field dependence from (2-3), 
including
the average supression of the gap magnitude in the vortex state, which we find to be
$\Delta_0(H)\simeq\Delta_0(0)[1-2\pi a^2H/(9H_{c2})]$.  The asymptotic expressions  
(5-6) are found to be quantitatively 
useful only up to fields of a few percent of $H_{c2}$ 
(above which terms linear in $H$ arising from contributions to $\kappa(\R)$ 
of order $\vs^4$ dominate) but reproduce the qualitative trends correctly
over a much larger scale.
 At higher temperatures the monotonic, quasilinear 
dependence on field is replaced
by a logarithmically decreasing function of field, as seen in the insert to the Figure.  
In the example shown, where impurity
scattering alone is acounted for, a minimum in $\kappa(H)$ is attained near $H/H_{c2}\simeq
 [T/(a\Delta_0)]^2$.  Inclusion of significant $e^-$--$e^-$ or $e^-$--vortex 
scattering will 
shift this minimum to higher fields.

{\it Scaling relations for $\kappa^{el}$.} While the application of the current theory to
the specific heat of a $d$-wave superconductor\cite{KH} indeed produces the exact, low energy
scaling function $F_C(T/\sqrt{H})$, the expression (6) may be
easily seen to violate the scaling property claimed for low-energy transport coefficients in a
$d$-wave superconductor\cite{SimonLee} due to the logarithms arising from the unitarity limit
relaxation rate.  While the authors of Ref. \cite{SimonLee} discussed
potential scattering,
they neglected multiple scattering thought to be important in HTSC.
\begin{figure}[h]
\begin{picture}(150,220)
\leavevmode\centering\includegraphics{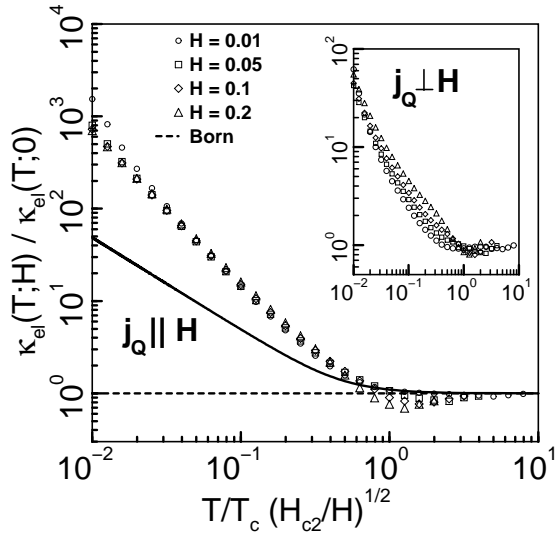}
\end{picture}
\caption{Scaling plot  $F_\kappa\equiv\kappa_el(T;H)/\kappa(T,H=0)$ 
vs. $X=(T/T_c)/(H/H_{c2})^{1/2}$ for ${\bf j}_Q\parallel {\bf H}$.
Symbols: constant field ``scans" in
unitarity limit showing 
approximate scaling: $H/H_{c2}=0.01,0.05,0.1,0.2$; solid line: constant
scattering; dashed line: Born limit. Insert: unitarity limit,  ${\bf j}_Q\perp
{\bf H}$.}
\end{figure}

Because the $T,H$ dependence of the logs is weak,
however, plotting $\kappa^{el}(T;H)/\kappa(T,H=0)$ vs. the scaling variable
$X\equiv T/\sqrt{H}$ will yield an {\it approximate} scaling in the unitarity limit.
For other types of scattering, exact scaling
functions   may be derived.  In  the (microscopically unjustified) 
case of constant
scattering rate ($1/\tau\sim$ const), 
one finds $F_\kappa(X)=F_C(X)$, where $F_C$ is the scaling function found
for the specific heat case,\cite{KopninVolovik2,KH}  whereas the 
(experimentally unrealistic) case of Born scattering yields a trivial
scaling, $F_\kappa=1$.
In Figure 2 we plot $F_\kappa$ for the three cases, illustrating how an
observation of scaling in a transport property can in principle yield
microscopic information about scattering processes.
Scaling should break down
completely either when 1) the impurity bandwidth
$\gamma$ becomes comparable to either
$E_H$ or
$T$, or 2) when the impurity relaxation rate becomes comparable to either the vortex
relaxation or inelastic collision rate.

{\it Dirty limit and inelastic scattering.}  As in the case of the specific heat, qualitatively
new behavior is obtained when the impurity bandwidth $\gamma\simeq 0.61 \sqrt{\Gamma\Delta_0}$ exceeds the
field scale $E_H$. At $T=0$ the variation with field has the same form as the clean limit given in Eq. (5),
but with $s=\pi^2\Delta_0^2/( 12\gamma^2)$ and $\rho=\sqrt{6 \gamma^2 H_{c2}/ (\pi \Delta_0^2 a^2 H)}$.
 At nonzero temperatures the minimum
in $\kappa(H)$ can be eliminated by impurities, as illustrated in Fig. 1.
We have also examined the effect of inelastic electron-electron scattering, and find 
that the rise in the conductivity due to field-induced quasiparticle states is eventually
compensated by the shortened mean free path, leading to a possible local maximum 
in $\kappa(H)$.

{\it Conclusions.} We have analyzed the contribution of  nodal quasiparticles 
 in the $d$-wave vortex state to magnetic field-dependent transport due to the
Doppler shifts of their energies in the flow field of the vortex lattice.  
We find an $H \log H$
increase at $T=0$, crossing over at $T>0$ to a possible minimum in  the 
field dependence of transport properties of clean  systems.  
If the mean free path due to scattering from vortex cores is
sufficiently long, {\it the field dependence is related directly
to the  temperature dependence in zero field}.  
This means that a theory of the zero-field $d$-wave state should be 
able to describe the field dependence with essentially no further parameters. 

Our results apply qualitatively to 3D order parameters with line nodes 
as well, and it is encouraging that data on thermal conductivity and 
scaling in $UPt_3$\cite{Flouquetscaling} appear to support our picture. 
The $T/H^{1/2}$ scaling  observed in this system may place an
important constraint on the order parameter symmetry, as no scaling is 
expected, e.g. for linear point nodes.  
We will discuss this case in  more detail elsewhere.  Data on field
dependence of thermal conductivity in HTSC currently available 
at $T \gtsim 5 K$ do not show the increase with
field predicted here.  We believe this is because inelastic scattering in these systems is
dominant down to a few Kelvin or even subKelvin temperatures, supressing the mean free path more
rapidly than the quasiparticle density of states is increased. Our understanding of the
nature of quasiparticles in the
$d$-wave state and potentially  of the effects seen in Refs.
\cite{Ong214,Ong2212} would be therefore aided by further measurements in these systems at lower
temperatures.

{\it Acknowledgements.}  The authors gratefully acknowledge 
important discussions with  J. Flouquet, A. Freimuth, J. Orenstein, T.V. Ramakrishnan, H. Suderow, L. Taillefer, and
P. W\"olfle.   Partial support was provided by
NSF-DMR-96--00105 and the A. v. Humboldt Foundation.  The authors are grateful
to the Institut f\"ur Theorie der Kondensierten Materie in Karlsruhe, where the
latter stages of the research were conducted.

\end{document}